\documentclass[aps,prl,onecolumn,reprint,showpacs,superscriptaddress,amsfonts,amsmath,amssymb]{revtex4-1}

\usepackage{graphicx}
\usepackage{color}
\definecolor{hblue}{rgb}{0.243,0.188,0.573}
\usepackage[colorlinks=true,citecolor=hblue,linkcolor=hblue,urlcolor=hblue,pdfborder={0 0 0}]{hyperref}
\usepackage{amsbsy}
\usepackage{xspace}
\usepackage{cleveref}
\usepackage[caption=false]{subfig}
\usepackage{ifthen}

\newcommand{\reprint}{0}

\ifthenelse{\equal{\reprint}{1}}{
    \newcommand{\figtype}{eps}
}{
	\usepackage{pdfpages}
	\newcommand{\figtype}{pdf}
	\pdfoutput=1
}

\renewcommand{\vec}[1]{\boldsymbol{#1}}
\newcommand{\vecr}{\ensuremath{\vec{r}}\xspace}
\newcommand{\Ef}{E_\text{F}}
\newcommand{\up}{\uparrow}
\newcommand{\down}{\downarrow}
\newcommand{\liqhe}{L^{\star}}
\newcommand{\dl}[1]{{\hat #1}}
\newcommand{\dlCdd}{\dl{C}_\text{dd}}
\newcommand{\dlEf}{\dl{E}_\text{F}}
\newcommand{\dlliqhe}{\dl{L}^{\star}}
\newcommand{\Vdd}{\ensuremath{V_\text{dd}}\xspace}
\newcommand{\Cdd}{\ensuremath{C_\text{dd}}\xspace}
\newcommand{\ef}[1]{\ensuremath{\operatorname{e}^{#1}}\xspace}

\newcommand{\absv}[1]{\left|#1\right|}
\newcommand{\absvsq}[1]{\absv{#1}^2}	
\newcommand{\ket}[1]{\left|#1\right\rangle}
\newcommand{\bra}[1]{\left\langle#1\right|}
\newcommand{\figref}[1]{Fig.~\ref{fig:#1}}
\newcommand{\bb}[1]{\left(#1\right)}

\begin{document}

\title{Driving Dipolar Fermions into the Quantum Hall Regime by\texorpdfstring{\\}{ } Spin-Flip Induced Insertion of Angular Momentum}

\author{David~Peter}
\email[Corresponding author: ]{peter@itp3.uni-stuttgart.de}
\affiliation{Institute for Theoretical Physics III, University of Stuttgart, Germany}
\author{Axel~Griesmaier}
\author{Tilman~Pfau}
\affiliation{5. Physikalisches Institut, University of Stuttgart, Germany}
\author{Hans~Peter~B\"{u}chler}
\affiliation{Institute for Theoretical Physics III, University of Stuttgart, Germany}

\date{\today}

\begin{abstract}
A new method to drive a system of neutral dipolar fermions into the lowest Landau level regime is proposed.
By employing adiabatic spin-flip processes in combination with a diabatic transfer, the fermions are pumped to higher orbital angular momentum states in a repeated scheme that allows for the precise control over the final angular momentum.
A simple analytical model is derived to quantify the transfer and compare the approach to rapidly rotating systems.
Numerical simulations of the transfer process have been performed for small, interacting systems.
\end{abstract}

\pacs{67.85.Fg, 67.85.Lm, 73.43.-f}
\maketitle

Despite being ideal models for complicated solid state systems, ultracold quantum gases lack one important aspect of the electronic complex: because of the charge neutrality of the atoms, there are no mobile charge carriers that possess a direct coupling to the magnetic vector potential. Plenty of interesting effects, however, arise when charged particles are subject to high magnetic fields in low dimensional systems. The most prominent ones are the integer quantum Hall effect \cite{Klitzing1980} as an example for the appearance of topological states, as well as the fractional quantum Hall effect \cite{Laughlin1983}, potentially giving rise to fundamental excitations with non-Abelian statistics.

Several schemes have been proposed to simulate the effect of magnetic fields for neutral particles. Artificial gauge fields can be created by imprinting phases, making use of the Peierls substitution in optical lattices \cite{Aidelsburger2011,Jimenez-Garcia2012,Struck2012}, or by tailoring spatially dependent Hamiltonians to generate geometric phases \cite{Lin2009}; for an overview see \cite{Dalibard2011}. Rapidly rotating quantum gases provide an alternative route via Larmor's theorem
\cite{Cooper2008,Fetter2009}. Several theoretical proposals demonstrate the appearance of highly correlated quantum Hall states for dipolar bosons \cite{Cooper2005} and fermions \cite{Baranov2005,Osterloh2007}. However, the experimental realization of quantum Hall states has been elusive so far. For rotating systems, the main problem is the precise control on the rotation frequency, which is required to reach the lowest Landau level without crossing the rotational instability \cite{Schweikhard2004}.

Here we propose a new scheme to access the regime of fast rotation for a dipolar Fermi gas such as $^{161}\text{Dy}$, which has recently been cooled to the quantum degenerate regime \cite{Lu2012}. Starting from a spin-polarized state, dipolar interactions can lead to spin relaxation with a net angular momentum transfer \cite{Hensler2003}. This is known as the Einstein--de Haas effect \cite{Einstein1915} and has been proposed to create rotating Bose-Einstein condensates \cite{Santos2006,Kawaguchi2006}. We suggest using this mechanism in a trapped, quasi-two-dimensional system to control the amount of angular momentum, and -- by repeated application of the transfer scheme -- reach the lowest Landau level (LLL). This scheme allows for direct control over the total angular momentum instead of the rotation frequency and circumvents the prime experimental difficulties toward the realization of the quantum Hall regime in harmonically trapped gases.


\begin{figure}[b]
	\subfloat[]{
		\includegraphics[height=0.32\columnwidth]{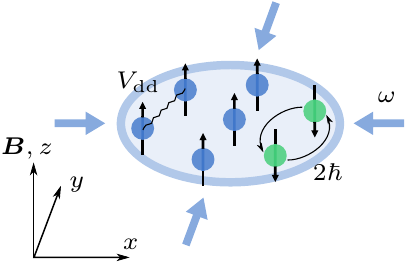}
		\label{fig:fig1a}
	}
	\subfloat[]{
		\includegraphics[height=0.32\columnwidth]{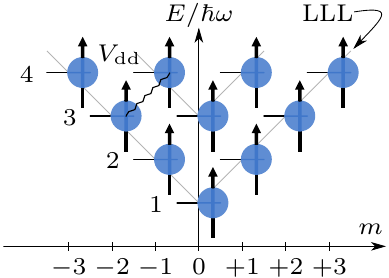}
		\label{fig:fig1b}
   }\\
	\subfloat[]{
		\includegraphics[height=0.32\columnwidth]{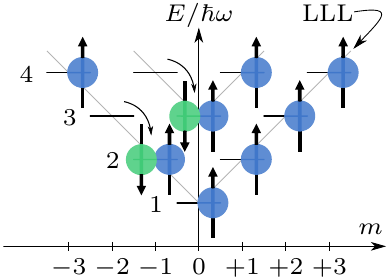}
		\label{fig:fig1c}
	}
	\subfloat[]{
		\includegraphics[height=0.32\columnwidth]{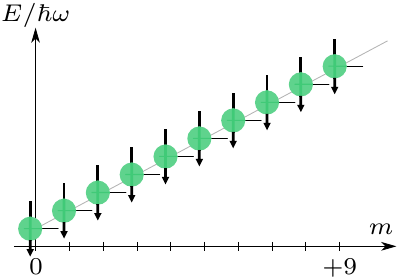}
		\label{fig:fig1d}
	}

  \caption{\label{fig:fig1}(a) Dipolar particles, trapped in a quasi-2D geometry with a radial confinement $\omega$. When the external magnetic field $\vec{B}$ is tuned in resonance, dipolar interactions \Vdd can induce spin relaxation processes, leading to a net angular momentum increase of $1\hbar$ per particle. (b) Energy levels of a 2D harmonic oscillator. (c) One of the possible spin-flip processes, bringing both particles to higher angular momentum states. (d) Eventually, after repeated application of the driving scheme, all particles occupy the lowest Landau level.}
\end{figure}

We consider a system of $N$ fermionic atoms with magnetic dipole moments $\vec{\mu}$. While extensions to schemes with polar molecules are possible, the permanent dipole moments of the atoms lead to some simplifications.
To shorten the discussion, we consider only two internal levels (pseudospin 1/2). The particles are confined in a quasi-two-dimensional harmonic trap with a radial frequency $\omega$ and an axial frequency $\omega_z$. For strong $z$ confinement $\hbar\omega_z \gg \Ef$, where $\Ef$ is the Fermi energy derived below, the system is effectively 2D, see~\figref{fig1a}.
The interactions between the particles are described by the dipolar interaction potential
\begin{align*}
\Vdd(\vecr) = \frac{\mu_0}{4\pi}\frac{\vec{\mu}_i  \vec{\mu}_j - 3(\vec{\mu}_i  \vec{\hat{r}})(\vec{\mu}_j \vec{\hat{r}})}{r^3}    
\end{align*}
where $\vec{r}=\vec{r}_i-\vec{r}_j$ is the relative distance between the two particles.
Note that a weak $s$-wave scattering length does not change the general behavior of our transfer scheme and is ignored in the following.
The dipole moment $\vec{\mu} =  \mu_\text{B}g \vec{S} / \hbar = \mu_\text{B}g {\boldsymbol\sigma}/2$ is given in terms of the Land\'e factor $g$ and the Pauli matrices. By integrating out the fast motion perpendicular to the $xy$ plane, taking the limit $\omega_z\rightarrow\infty$, and using the spin raising and lowering operators $\sigma^{\pm} = (\sigma^x \pm i \sigma^y)/2$ the interaction reduces to
\begin{align*}
\Vdd(r,\phi) =  \frac{\Cdd}{r^3} \left[ 
\sigma^z_i \sigma^z_j - (\sigma^+_i \sigma^-_j +3 \ef{2i\phi} \sigma^-_i \sigma^-_j  +\text{h.c.})
\right]
\end{align*}
where $r,\phi$ are polar coordinates in the $xy$ plane and $\Cdd = \mu_0 \mu_\text{B}^2 g^2 / 16\pi$  characterizes the strength of the interaction. The dipolar interaction features three different processes. The first term proportional to $\sigma^z_i\sigma^z_j$ describes spin-preserving collisions, while the second term \smash{$\sigma^+_i\sigma^-_j$} accounts for spin-exchange collisions. These terms conserve separately the total spin and the total angular momentum.  Finally, the third operator \smash{$\ef{2i\phi} \sigma^-_i \sigma^-_j$} describes the relaxation process that transfers spin to orbital angular momentum, see \figref{fig1b},\subref*{fig:fig1c}. The sum $L+S$ is still conserved and the spin flip leads to an orbital motion with an increase of relative angular momentum of $2\hbar$.

It is this process that allows us to drive the dipolar particles to higher angular momentum states. Assuming the gas is initially in a spin-polarized state with the external magnetic field pointing in the positive $z$ direction, the particles will undergo spin relaxation when the field is adiabatically ramped through zero and finally pointing in the negative $z$ direction. During this adiabatic ramping, 
the total orbital angular momentum is increased by $N\hbar$ with $N$ the number of particles in the system.  For the goal to reach the lowest Landau level regime, it is required to transfer  $\liqhe\equiv N(N-1)/2 \cdot \hbar$ angular momentum to the orbital degrees of freedom, as described below. It is therefore necessary to reverse the magnetic field and the spins to their original position, in a way that guarantees repeated application of the transfer scheme without affecting the orbital angular momentum.

To achieve this, we propose rotating the magnetic field by $180^\circ$ around an arbitrary axis lying in the $xy$ plane (say, the $y$ axis), slowly enough such that the spins rotate adiabatically, but fast enough such that the orbital degrees of freedom cannot follow. To satisfy the adiabaticity with respect to the spins and diabaticity with respect to the external degrees of freedom, the speed of the rotation $\gamma_\text{rot}$ has to satisfy
$\omega \ll \gamma_\text{rot} \ll \omega_\text{L}$, where $\omega_\text{L}=g\mu_B B/\hbar$ is the Larmor frequency. After the rotation, the magnetic field has enclosed a \textsf{D}-shaped path in the $xz$ plane. The spins are now pointing upward (in analogy to \figref{fig1b} but with
increased angular momentum) and the transfer scheme can be applied again. Multiple repetitions are realistic and only limited by the finite lifetime of the trapped ensemble.

High angular momentum states are indeed related to the quantum Hall regime, as there is a close connection between the Landau levels and the states $\ket{n,m}$ of a two-dimensional harmonic oscillator in terms of a radial quantum number $n=0,1,\dots$ and angular momentum $\hbar m$, see \figref{fig1b}. In particular, the ground state of $N$ fermions filled into the harmonic oscillator with the constraint $L=\liqhe$ is given by the many-body state
\begin{align*}
\Psi&= \bra{\{z_i\}}\mathcal{A} \prod_{m=0}^{N-1} \!\ket{0,m} = \mathcal{N}\Bigg[\prod_{i<j} (z_i-z_j) \Bigg] \ef{-\frac{1}{2}\sum\absvsq{z_k}}.
\end{align*}
Here $z_k=(x_k+i y_k)/l_\text{HO}$ are complex coordinates of the particles, $\mathcal{A}$ is the antisymmetrizer, $\mathcal{N}$ is a normalization constant, and \smash{$l_\text{HO}=\sqrt{\hbar/m\omega}$} is the harmonic oscillator length. This wave function is equivalent to the Laughlin wave function for integer filling $\nu=1$, with   \smash{$l_\text{HO}=\sqrt{\hbar/m\omega}$} replacing the magnetic length \smash{$\sqrt{2}l_m=\sqrt{2\hbar c/eB}$} for electronic systems. Quite generally, the states with $n=0$ and $m\ge 0$ correspond to the states in the lowest Landau level, see \figref{fig1d}.
To reach the LLL regime, we have to repeat the transfer scheme at least $\liqhe/N\hbar=(N-1)/2$ times.

\begin{figure}[t]
	\includegraphics[width=0.8\columnwidth]{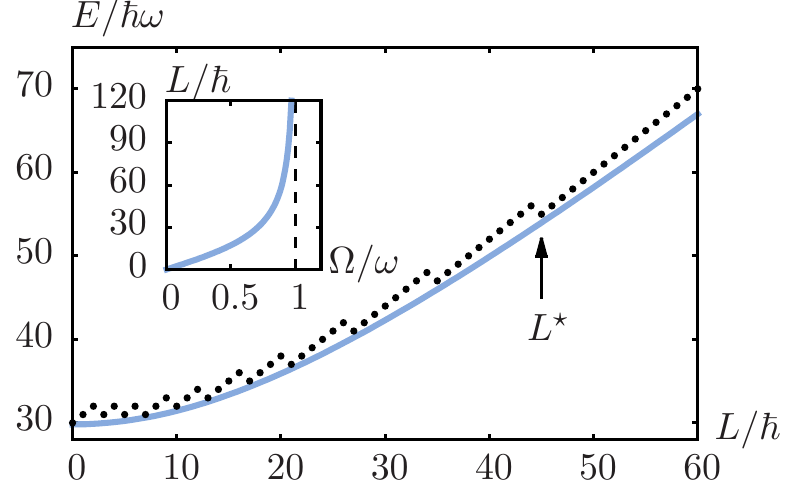}

    \caption{\label{fig:fig2} Exact ground state energy (dots) for $N=10$ particles at fixed angular momentum $L$, compared to the approximate expression (solid line) as given in~\eqref{eq:etotnl}. For $L>\liqhe=45\hbar$, the energy increases linearly. Inset shows $L$ as a function of the rotation frequency $\Omega$ in the analytic model. $L$ diverges at the critical rotation frequency $\Omega=\omega$, when the rotation exceeds the trap frequency.}
\end{figure}

To quantify a single transfer process, our first aim is to calculate the total energy of $N$ harmonically trapped fermions for a fixed total angular momentum $L$ (polarized state, one spin component). For the noninteracting system, the energy can be obtained by simple summations. We start with the ground state for $L=0$, where all energy shells up to the Fermi energy are completely filled. The energy of the single particle states~$\ket{n,m}$ is given by $E_{nm}=\hbar\omega(2n+\absv{m}+1)$. To avoid cluttering of notation, we introduce dimensionless quantities indicated by a $\dl{\ }$ sign. These quantities are measured in oscillatory units. That is, energy in units of $\hbar\omega$, angular momentum in units of $\hbar$, lengths in units of $l_\text{HO}$ and time in units of $\omega^{-1}$. The degeneracy of each energy level is simply given by $g(\dl{E}) = \dl{E}$.
With $N=\sum g(\dl{E}) = \dlEf(\dlEf+1)/2$ the Fermi energy is determined by
\begin{align*}
\dlEf &= \frac{1}{2}\bb{\sqrt{8N+1}-1} \xrightarrow{N\gg 1} \sqrt{2N}
\end{align*}
The total energy for $N$ particles is then given by
\begin{align} \label{eq:energyzero}
\dl{E}(N) = \sum_{\dl{E}=1}^{\dlEf} g(\dl{E}) \dl{E} = \frac{N}{3} \sqrt{8N+1} \xrightarrow{N\gg 1} \frac{(2N)^{3/2}}{3}
\end{align}
which shows the known scaling of a trapped 2D Fermi gas~\cite{Yoshimoto2003}. 
To derive the total energy $E(N,L)$ for $L\ne 0$, we define $N_m$ as the number of particles with angular momentum $m$. The energy in terms of $N_m$ is given by
\begin{align} \label{eq:energyf}
\dl{E} = \sum_m \sum_{n=0}^{N_m-1} \dl{E}_{nm} = \sum_m N_m\left(N_m + \absv{m} \right).
\end{align}
The exact ground state energy can be found combinatorially for small particle numbers by varying the $N_m$ for fixed $N$ and $L$. The result for $N=10$ is shown in~\figref{fig2}. For larger particle numbers this method is not feasible, but an analytic solution can be found for large particle numbers. Then, we can treat $N_{m}$ as a continuous function. To find the minimum of \eqref{eq:energyf} at fixed $N$ and $L$, we introduce two Lagrange multipliers $\mu$, $\Omega$ for the conditions $N = \sum_m N_m$ and $\dl{L} = \sum_m N_m m$, respectively. Taking the functional derivative with respect to $N_m$ yields
$N_m=(\dl{\mu}-\absv{m}+\dl{\Omega} m)/2$.
The parameters can be determined by solving the constraints and summing from $m_-=-\dl{\mu}/(1+\dl{\Omega})$ to $m_+=\dl{\mu}/(1-\dl{\Omega})$, where $N_{m_\pm}=0$. One finds
\begin{align*}
\dl{\Omega} = \frac{3 \dl{L}}{\sqrt{\big(2N\big)^3 + \big(3\dl{L}\big)^2}}, \qquad \dl{\mu}= \frac{N^2}{\sqrt{\big(2N\big)^3 + \big(3\dl{L}\big)^2}}.
\end{align*}
By using these relations and omitting correction terms of order $1/L$ and $\sqrt{N}$, we obtain the total energy
\begin{align} \label{eq:etotnl}
\dl{E}(N,\dl{L})= \frac{1}{3} \sqrt{\big(2N\big)^3 + \big(3\dl{L}\big)^2}.
\end{align}
This result agrees with the exact behavior for $L=0$ as derived in~\eqref{eq:energyzero}, and even for particle numbers as small as $N=10$ it is close to the exact ground state energy, as shown in \figref{fig2}. For $L\ge\liqhe$, the minimization problem becomes trivial as all particles occupy the lowest Landau level. The energy is exactly given by $\dl{E}=\dl{L}$, which is also obtained asymptotically from~\eqref{eq:etotnl} in the limit $\dl{L}\gg N$.

It is now possible to quantify the link between our approach (fixed angular momentum) and rapidly rotating systems (fixed rotation frequency) explicitly. Both are connected by a Legendre transform and we should in fact interpret the Lagrange multiplier $\Omega=\frac{\partial E}{\partial L}$
as the rotation frequency. In a harmonic trap, the system becomes unstable if $\Omega$ exceeds the value of the trap frequency $\omega$, as the harmonic confinement in the rotating frame is effectively given by $\omega-\Omega$. The angular momentum
\begin{align*}
\dl{L}=\frac{(2N)^{3/2}}{3} \frac{\dl{\Omega}}{\sqrt{1-\dl{\Omega}^2}}
\end{align*}
has a singularity for $\dl{\Omega}=\Omega/\omega=1$ and large values of $L$ can only be achieved by tuning $\Omega$ close to the critical value. It is this precise control on the rotation frequency that so far prevented the experimental realization of the quantum Hall regime in harmonically trapped gases. In contrast, for the present situation, the system is always stable as $\Omega(L)<\omega$ for all $L$. An arbitrary orbital angular momentum can be transferred to the system by the ramping scheme with high precision.


\begin{figure*}[t]
	\subfloat[]{
		\includegraphics[height=0.4\columnwidth]{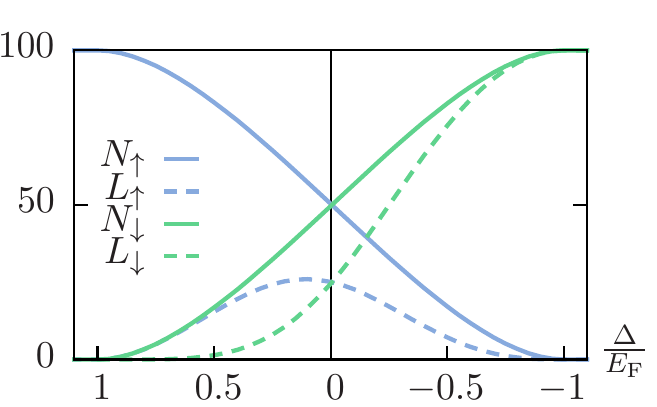}
		\label{fig:fig3a}
	}
	\subfloat[]{
		\includegraphics[height=0.4\columnwidth]{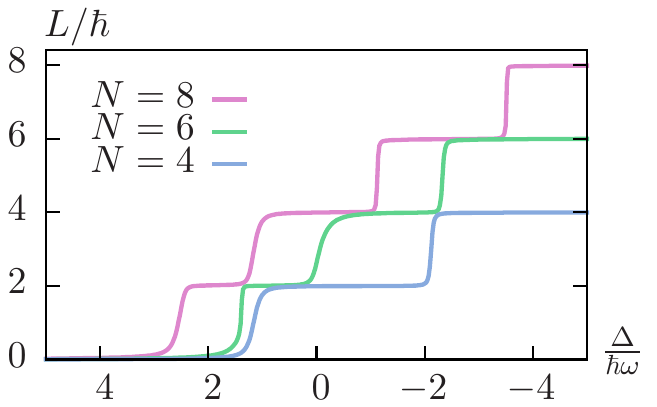}
		\label{fig:fig3b}
	}
	\subfloat[]{
		\includegraphics[height=0.4\columnwidth]{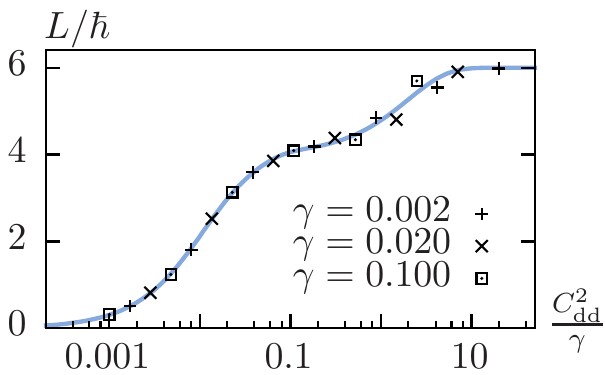}
		\label{fig:fig3c}
	}
  	\caption{\label{fig:fig3}(a) Description of the transfer in the analytical model with $N=100$ particles for decreasing energy splitting $\Delta$ between the two components $\up$ and $\down$. The transfer starts at $\Delta=\Ef$ with particles continuously being transferred into the $\down$ state as $\Delta$ is lowered to $-\Ef$. Notice that during the transfer, both components rotate in the same direction. The crossing $N_\up=N_\down$ is not precisely at $\Delta=0$ due to the initial bias. (b) Full simulation of the transfer scheme for $N=4,6$, and $8$~particles in the adiabatic limit $\gamma\rightarrow 0$. As the Zeeman splitting $\Delta$ is tuned through zero, the angular momentum increases in steps of $2\hbar$, indicating the transfer of two particle at a time. The interaction strength is given by $\dlCdd=0.1$. (c) Angular momentum at the end of the transfer for $N=6$ particles at different values of the Landau-Zener parameter $\lambda=\dlCdd^2/\dl{\gamma}$. The data points for different rates collapse onto a single curve. The solid line is a probabilistic model, fitted to the data points.}
\end{figure*}

Starting from expression \eqref{eq:etotnl} for the energy, we are now able to describe the transfer process in the adiabatic limit. Let $N_\up$ be the number of particles in the spin-up state and $N_\down = N-N_\up$ the particles in the spin-down state. We describe both components separately and write the total energy as
$E(N_\up,L_\up) + E(N_\down,L_\down) + \Delta\cdot N_\down$
where we have introduced the Zeeman energy shift $\Delta=\mu_\text{B} g B$ (energy measured with respect to the energy of the lower Zeeman state). We assume that every particle eventually takes part in the transfer process (adiabaticity) and consequently one quanta of angular momentum is transferred per particle. Starting from the nonrotating state at $L=0$, this imposes the transfer condition $L_\up + L_\down = L = N_\down\hbar$. Adding this condition with another Lagrange multiplier, one can quantify the transfer process as a function of $\Delta$, see \figref{fig3a}. Coming from high fields where $\Delta > \Ef$, the transfer starts right at the Fermi energy. Note that during the transfer, while $\Ef > \Delta > -\Ef$, both components ($\up$, $\down$) rotate in the same direction. Eventually all particles get transferred to the lower spin state and the total angular momentum equals $L=L_\down=N\hbar$.

To justify the adiabaticity assumption above, we simulate the transfer process for small systems of few particles. We include all interactions mediated by $\Vdd(r,\phi)$, and assume, that the strength of the interaction $\dlCdd=(\Cdd/l_\text{HO}^3)/\hbar\omega \ll 1$ is weak compared to the Landau level splitting. Then, only a few excited states have to be taken into account.
The system dynamics is described by
\begin{align*}
H=\sum_{i} \left[ E_{nm}  + \Delta(t)\,\delta_{\sigma,\down} \right] {c}^\dagger_{i}{c}^{\phantom\dagger}_{i} +  \frac{1}{2}\sum_{ijkl} V_{ijkl}\, {c}^\dagger_{i} {c}^\dagger_{j} {c}^{\phantom\dagger}_{l}{c}^{\phantom\dagger}_{k}
\end{align*}
where each of the indices $ijkl$ of the fermionic operators labels a set of quantum numbers $(n,m,\sigma)$ and $\Delta(t)/\hbar\omega = -\gamma t$ is the time-dependent Zeeman shift, controlled by the linearly decreasing magnetic field. The calculation of the dipolar interaction matrix elements $V_{ijkl}\sim\Cdd$ is presented in the supplemental material. The only relevant parameters in this model are the transfer rate  $\dl{\gamma}=\gamma/\omega$ and the interaction strength $\dlCdd$.
For the perfect adiabatic transfer, in the limit $\gamma\rightarrow 0$, we can find the instantaneous ground state of $H$ as $\Delta$ decreases. The results are shown in \figref{fig3b} for $N=4,6$, and $8$ particles. The total angular momentum $L(\Delta)$ increases gradually from $L=0$ to $L=N\hbar$ in steps of $2\hbar$, indicating that two particles are transferred at a time.

To obtain results for a finite transfer rate $\gamma$, we simulate the full time-dependent many-body problem.
The total angular momentum $L(t\rightarrow\infty)$ at the end of the transfer for $N=6$ particles is shown in \figref{fig3c} for different values of $\dlCdd$ and $\dl{\gamma}$.  Remarkably, the data points collapse onto a single line using  $\lambda = \dlCdd^2/\dl{\gamma}$. This parameter arises in the Landau-Zener formula of a single level crossing, and the collapse indicates that 
each pair transfer is dominated by an individual avoided level crossing. A simple model accounting for this behavior (solid line) describes the final angular momentum observed in the full simulation (see supplemental material).

The preparation of the integer quantum Hall state with an orbital angular momentum of $\dlliqhe = N(N-1)/2$ is finally achieved by a sequence of ramping cycles: Starting with an unpolarized sample with the fermions equally distributed between the two spin states, i.e.,  $N_\up=N_\down=N/2$, a first transfer increases the orbital angular momentum by only $\dl{L}=N/2$. Then, $N/2-1$ subsequent cycles will transfer exactly the required orbital angular momentum to reach the integer quantum Hall state. 

In an experimental realization with $^{161}\text{Dy}$ atoms, the number of cycles can be significantly reduced due to the total spin of $F=21/2$ in the hyperfine ground state. Although calculations for $22$ internal levels are too complex, we expect no qualitative modifications, except that $21\hbar$ of angular momentum are transferred per particle and cycle \footnote{Other highly dipolar fermions used in cold atom experiments are $^{167}\text{Er}$ and $^{53}\text{Cr}$ with a total angular momentum of $F=19/2$ and $9/2$, respectively \cite{Berglund2007,Chicireanu2006}. They could therefore provide $19\hbar$ or $9\hbar$ of angular momentum per atom and transfer.}. Two important experimental requirements are a precise magnetic field control \cite{Pasquiou2011} as well as a deterministic preparation scheme for a certain particle number, as demonstrated in \cite{Serwane2011}. For the magnetic field ramp we can estimate an optimal minimum value for the rate $\dl{\gamma}=2\dlEf/\dl{t}_\text{e}=2\sqrt{2N}/\omega t_\text{e}$ by observing that the Zeeman splitting has to be tuned at least once from $\Ef$ to $-\Ef$ within the experimental accessible time $t_\text{e}$, which is limited by the lifetime of the atoms in the trap. The Landau-Zener parameter is finally given by
\begin{align*}
\lambda = \frac{\omega t_\text{e}}{2\sqrt{2N}} \left(\frac{l_\text{DDI}}{4 l_\text{HO}}\right)^2
\end{align*}
where the length $l_\text{DDI} = m\mu_0\mu^2/4\pi \hbar^2$ parametrizes the strength of the interaction \cite{Lu2012}. In a setup with $N\sim 10$ fermionic $^{161}\text{Dy}$ atoms, a long lifetime of $t_\text{e}=10\text{s}$ and a radial frequency of $\omega = 3\text{kHz}$ are needed to reach values of $\lambda$ on the order of $1$. We comment, however, that the transfer scheme works already for smaller values of $\lambda$.


A particularly interesting property of the integer quantum Hall state, potentially useful to detect the successful generation, is the perfectly flat density $n=1/\pi l_\text{HO}^2$ within a circular region of radial size $\sqrt{N}l_\text{HO}$.
In addition, it is possible to reach states with $L>\liqhe$ by continuing the transfer scheme.
In this regime,
highly correlated ground states appear that are closely connected to fractional quantum Hall states, see~\cite{Osterloh2007} for a discussion in the context of rotating systems. Consequently, the presented method of dipolar relaxation allows for the exploration of integer and fractional quantum Hall states, but avoids the experimentally challenging requirement of precise control of the rotation frequency by directly tuning the orbital angular momentum.

\begin{acknowledgments}
We acknowledge the support from the Deutsche For\-schungs\-gemeinschaft (DFG) within the SFB/TRR~21.
\end{acknowledgments}

%

\ifthenelse{\equal{\reprint}{1}}{}{
	\clearpage
	

\section*{Supplementary material}

David Peter,[1] Axel Griesmaier,[2] Tilman Pfau,[2] and Hans Peter Büchler[1]

[1] Institute for Theoretical Physics III, University of Stuttgart, Germany
[2] 5. Physikalisches Institut, University of Stuttgart, Germany

**Matrix elements of the dipolar interaction**

In this section, we show how to simplify the matrix elements of the dipolar interaction $V_{ijkl} = \langle ij|V_{\rm dd}|kl\rangle$ with each index $ijkl$ representing a set of quantum numbers $i = (n_i, m_i, \sigma_i)$ and

$$V_{\rm dd}(r,\phi) = \frac{C_{\rm dd}}{r^3} \left[\sigma_1^z \sigma_2^z - (\sigma_1^+ \sigma_2^- + 3\,e^{2i\phi}\,\sigma_1^- \sigma_2^- + \text{h.c.})\right]$$

the dipolar interaction in 2D in terms of the polar coordinates of the relative vector $\boldsymbol{r} = \boldsymbol{r_1} - \boldsymbol{r_2}$ between two particles. The spin-part is easily resolved and therefore we can concentrate on matrix elements of the form

$$V^{\Delta m} \equiv \left\langle n_1' m_1' n_2' m_2' \left| \frac{e^{i\Delta m \phi}}{r^3} \right| n_1 m_1 n_2 m_2 \right\rangle$$

with $\Delta m = 0, \pm 2$. It is useful to change the basis to center-of-mass and relative coordinate states with

$$\boldsymbol{Q} = (\boldsymbol{r_1} + \boldsymbol{r_2})/\sqrt{2}, \qquad \boldsymbol{q} = (\boldsymbol{r_1} - \boldsymbol{r_2})/\sqrt{2}.$$

Notice the symmetric definition with the additional factor of $1/\sqrt{2}$ compared to the usual definition of the relative vector $\boldsymbol{r} = \sqrt{2}\boldsymbol{q}$. Due to the quadratic character of the potential, $\boldsymbol{Q}$ and $\boldsymbol{q}$ are subject to the same harmonic potential. Thus, the product state $|n_1 m_1 n_2 m_2\rangle = |n_1 m_1\rangle |n_2 m_2\rangle$ can be decomposed in terms of harmonic oscillator states $|NM\rangle|nm\rangle$ of the $\boldsymbol{Q}, \boldsymbol{q}$ coordinates via

$$|n_1 m_1 n_2 m_2\rangle = \sum_{N,M,n,m} T^{n_1 m_1 n_2 m_2}_{NMnm} |NMnm\rangle$$

where the $T^{n_1 m_1 n_2 m_2}_{NMnm}$ are called Talmi-Moshinsky coefficients [1, 2]. Since the center-of-mass is not affected by the interaction, the relevant matrix elements are

$$\left\langle n'm' \left| \frac{e^{i\Delta m \phi}}{(\sqrt{2}q)^3} \right| nm \right\rangle = \delta_{m+\Delta m, m'} \int_0^\infty \mathrm{d}q\, \frac{R_{n'}^{m'}(q) R_n^m(q)\, q}{(\sqrt{2}q)^3}$$

where the radial functions $R_n^m$ are given in terms of the generalized Laguerre polynomials $L_n^{|m|}$ as

$$R_n^m(q) = \sqrt{\frac{2n!}{(n+|m|)!}}\, q^{|m|} \exp\left(-\frac{q^2}{2}\right) \cdot L_n^{|m|}(q^2).$$

The matrix element $V^{\Delta m}$ is thus

$$V^{\Delta m} = \sum_{N,M,n,n',m} (T^*)^{n_1' m_1' n_2' m_2'}_{NMn'(m+\Delta m)} \cdot T^{n_1 m_1 n_2 m_2}_{NMnm} \cdot$$
$$\cdot \int_0^\infty \mathrm{d}q\, \frac{R_{n'}^{m+\Delta m}(q) R_n^m(q)\, q}{(\sqrt{2}q)^3}.$$

where the remaining integral can be calculated analytically for specific $n, n', m$ and $\Delta m$.

**Probabilistic model for $L_N$**

Here we derive the total amount of angular momentum after the transfer by assuming that each 2-particle process is described by an independent Landau-Zener (avoided) crossing. That is, we neglect any interference effects. For each Landau-Zener process, we define the probability to transfer the $n$-th pair of particles by $P_n = 1 - e^{-\lambda/\lambda_n}$ with $\lambda = \hat{C}_{\rm dd}^2/\hat{\gamma}$ the Landau-Zener parameter and $\lambda_n$ an effective coupling strength, describing the $n$-th pair-transfer process. The total angular momentum for $N$ particles after one cycle is then given by weighting each possible outcome ($\hat{L}=0, \hat{L}=2, \ldots, \hat{L}=N$) by the respective probability

$$\hat{L}_N = \sum_{n=1}^{N/2-1} 2n\, P_1 \cdots P_n (1 - P_{n+1}) + N P_1 \cdots P_{N/2}$$
$$= 2P_1 \left(1 + P_2 \left(1 + P_3 \left(1 + \ldots \left(1 + P_{N/2}\right)\right)\right)\right)$$

For $N=6$ particles this reduces to

$$\hat{L}_6 = 2P_1(1 + P_2(1 + P_3))$$
$$= 2\left(1 - e^{-\lambda/\lambda_1}\right)\left(1 + \left(1 - e^{-\lambda/\lambda_2}\right)\left(2 - e^{-\lambda/\lambda_3}\right)\right)$$

The assumption of independent crossings can now be justified a-posteriori. By fitting $\hat{L}_6$ to the simulation data we find $\lambda_1 = 0.0056$, $\lambda_2 = 0.025$, $\lambda_3 = 1.74$ with $\lambda_1 \ll \lambda_2 \ll \lambda_3$. While we suspect this approximation to break down for larger $N$, the model describes the transfer for small particle numbers in good agreement with the simulation.


}

\end{document}